\documentclass[10pt,letterpaper]{article}
\usepackage[top=0.85in,left=2.75in,footskip=0.75in]{geometry}

\usepackage{amsmath,amssymb}

\usepackage{changepage}

\usepackage[utf8x]{inputenc}

\usepackage{textcomp,marvosym}

\usepackage{cite}

\usepackage{nameref,hyperref}

\usepackage[right]{lineno}

\usepackage{microtype}
\DisableLigatures[f]{encoding = *, family = * }

\usepackage[table]{xcolor}

\usepackage{array}

\newcolumntype{+}{!{\vrule width 2pt}}

\newlength\savedwidth

\newcommand\thickhline{\noalign{\global\savedwidth\arrayrulewidth\global\arrayrulewidth 2pt}%
\hline
\noalign{\global\arrayrulewidth\savedwidth}}

\usepackage{setspace} 
\raggedright
\usepackage[aboveskip=1pt,labelfont=bf,labelsep=period,justification=raggedright,singlelinecheck=off]{caption}

\makeatletter
\renewcommand{\@biblabel}[1]{\quad#1.}
\makeatother

\usepackage{lastpage,fancyhdr,graphicx}
\usepackage{epstopdf}
\fancyheadoffset[L]{2.25in}
\fancyfootoffset[L]{2.25in}
\usepackage{graphicx}

\begin{document}
\vspace*{0.2in}

\begin{flushleft}
{\Large
\textbf\newline{A city of cities: Measuring how 15-minutes urban accessibility shapes human mobility in Barcelona} % Please use "sentence case" for title and headings (capitalize only the first word in a title (or heading), the first word in a subtitle (or subheading), and any proper nouns).
}
\newline
\\
Eduardo Graells-Garrido\textsuperscript{1,4,*},
Feliu Serra-Burriel\textsuperscript{1,2},
Francisco Rowe\textsuperscript{3},
Fernando M. Cucchietti\textsuperscript{1},
Patricio Reyes\textsuperscript{1}
%Name6 Surname\textsuperscript{2\ddag},
%Name7 Surname\textsuperscript{1,2,3*},
%with the Lorem Ipsum Consortium\textsuperscript{\textpilcrow}
\\
\bigskip
\textbf{1} Barcelona Supercomputing Center (BSC), Barcelona, Catalonia, Spain
\\
\textbf{2} Department of Statistics and Operations Research, Universitat Polit\`{e}cnica de Catalunya, Barcelona, Catalonia, Spain
\\
\textbf{3} Geographic Data Science Lab, Department of Geography and Planning, University of Liverpool, Liverpool, United Kingdom
\\
\textbf{4} Data Science Institute, Universidad del Desarrollo, Santiago, Chile
\bigskip

% Insert additional author notes using the symbols described below. Insert symbol callouts after author names as necessary.
% 
% Remove or comment out the author notes below if they aren't used.
%
% Primary Equal Contribution Note
%\Yinyang These authors contributed equally to this work.

% Additional Equal Contribution Note
% Also use this double-dagger symbol for special authorship notes, such as senior authorship.
%\ddag These authors also contributed equally to this work.

%\textcurrency Current Address: Dept/Program/Center, Institution Name, City, State, Country % change symbol to "\textcurrency a" if more than one current address note
% \textcurrency b Insert second current address 
% \textcurrency c Insert third current address


% Group/Consortium Author Note
%\textpilcrow Membership list can be found in the Acknowledgments section.

% Use the asterisk to denote corresponding authorship and provide email address in note below.
* eduardo.graells@bsc.es

\end{flushleft}
\section*{Abstract}
As cities expand, human mobility has become a central focus of urban planning and policy making to make cities more inclusive and sustainable. Initiatives such as the ``15-minutes city'' have been put in place to shift the attention from monocentric city configurations to polycentric structures, increasing the availability and diversity of local urban amenities. Ultimately they expect to increase local walkability and increase mobility within residential areas.
While we know how urban amenities influence human mobility at the city level, little is known about spatial variations in this relationship.
Here, we use mobile phone, census, and volunteered geographical data to measure geographic variations in the relationship between origin-destination flows and local urban accessibility in Barcelona. 
Using a Negative Binomial Geographically Weighted Regression model, we show that, globally, people tend to visit neighborhoods with better access to education and retail. Locally, these and other features change in sign and magnitude through the different neighborhoods of the city in ways that are not explained by administrative boundaries, and that provide deeper insights regarding urban characteristics such as rental prices.
In conclusion, our work suggests that the qualities of a 15-minutes city can be measured at scale, delivering actionable insights on the polycentric structure of cities, and how people use and access this structure.

%\linenumbers

% Use "Eq" instead of "Equation" for equation citations.
\section*{Introduction}

Rates of urbanization have been rapidly increasing across all world's regions over the last century~\cite{united20182018}. Currently more than half of the global population live in urban areas and increasingly in highly-dense cities~\cite{united20182018}. Cities are projected to absorb more than two-thirds of the growth in global population, with 68 percent of the world’s population living in urban areas by 2050~\cite{desa2019world}. Making cities inclusive, safe, resilient, and sustainable has thus become a global priority~\cite{griggs2013sustainable}. 

Human mobility within cities is recognised as a key area of policy intervention to achieve these goals. Prior research has established a connection between mobility within cities and urban form~\cite{gordon1989influence, muniz2005urban}. As cities expand, average travel time and distance have increased giving rise to a debate regarding the optimal city size and structure~\cite{ewing2015compactness, kirkley2018betweenness}.  Urban sprawl is considered to be detrimental to city livability and sustainability, inducing longer within-city travel~\cite{batty2003traffic}, increased private car usage~\cite{brueckner2000urban, glaeser2004sprawl}, congestion~\cite{bento2005effects}, obesity~\cite{ewing2003relationship}, and pollution~\cite{anderson1996urban}. We also know that longer commutes are more prevalent in certain populations, particularly highly educated, affluent, male and dual-earner populations~\cite{green1999longer, white1986sex}. Yet distinctive patterns exist in less developed and unequal societies in which the poorest segments of the population tend to commute longer distances~\cite{lucas2012transport, lucas2016transport}. 

To promote more localised mobility patterns, current planning strategies have used smart growth interventions to move away from monocentric city structures to polycentric configurations~\cite{brezzi2015assessing, veneri2015urban}, with ideas such as a 15-minutes city gaining salience in policy circles. In essence, interventions promoting polycentricity seek to increase the offer of local amenities, such as schools, public transport options, health care facilities, food outlets, jobs, recreational areas, and retail shops, creating sustainable, inclusive, and walkable local areas within a small radius~\cite{veneri2015urban, artmann2019smart}. The exploration of local accessibility, regional accessibility and transit access opportunities is of utmost importance for the improvement of livability in cities. Previous studies suggest the need to focus on planning for pedestrian-scale neighborhoods to enhance local accessibility, adapt travel behaviors, and reduce car ownership and vehicle miles traveled~\cite{krizek2003residential, Zhang2020}.
Existing research indicates that spatial concentration of job, public services and commercial activity influences human mobility in cities, giving rise to destination hotspots or multiple centres~\cite{roth2011structure, louail2014mobile, bassolas2019hierarchical}. Yet, less is known about how much the availability and diversity of local amenities delivers the goal of retaining local population locally, and the ways in which the relationship between local amenities and mobility intensity varies geographically and interacts with the socio-demographic profile of resident populations. 

To address these gaps, we aim to model how local amenities contribute to origin-destination flows of the  resident population in a city; to quantify the extent of spatial variability in this influence; and, to assess emerging patterns associated with the socio-demographic profile of local populations in Barcelona, Spain. Traditionally, surveys and travel diaries are used to solve these tasks. While valuable, these data sources are expensive, infrequent, offer low population coverage, lack statistical validity at fine geographical scales, and become available with a lag of one or two years after collection~\cite{hidalgo2008dynamics, tao2018travel}. Mobile phone data has emerged as a novel source to capture human mobility~\cite{gonzalez2008understanding}. While they present shortcomings, mobile phone data are generated at unprecedented temporal frequency and geographical granularity. They record data for the selected population of mobile phone users and raise concerns of personal privacy and confidentiality. Ensuring anonymization, mobile phone data enable developing a more refined spatio-temporal and contemporary understanding of human mobility~\cite{de2013unique}. 

A collaboration between public and private entities, the Barcelona city council and Vodafone, has enabled us to access aggregated mobile phone data of Barcelona to capture its local patterns of human mobility. Barcelona provides a unique study area to investigate the geographical variability in the way urban amenities influence local patterns of mobility. Over the last two decades, and more recently pushed by the COVID-19 pandemic, various initiatives have been put in motion to deliver the concept of self-sufficient neighborhoods, or the 15-minutes city, in which local areas have the required offer of urban amenities to be accessed within a short distance in a residential area (one example of this is the distribution of food markets in the city~\cite{provansal1992els}). Yet, little empirical evidence exists as to the extent that local urban amenities reduce the intensity and distance of mobility. We draw from volunteered geographical information data from OpenStreetMap (OSM) to measure the availability and accessibility to local urban amenities or Points of Interest (POI), jointly with data from the Spanish population census to characterize the socio-demographic composition of local populations. Then, to assess the influence of urban amenities on mobility patterns, we first use a Negative Binomial (NB) regression model to identify the key system-wide urban amenities relating to low mobility intensity (or high retention rates) and how mobility patterns vary across population groups. Second, we apply a NB Geographically Weighted Regression (GWR) model to assess the extent and direction of spatial variation in these relationships. The result of this analysis is a set of vectors that characterizes the accessibility and mobility of each neighborhood, which may be different from the global city patterns. We also demonstrate how these vectors open further local analyses, by studying their relationship with rental prices in the city.

The paper is structured around three main sections. First, we provide a description of the data sources to capture mobility, accessibility to urban amenities and characterize the socio-demographic composition of the local resident population. We also discuss the statistical modeling approach employed to measure the global and local relationship between urban amenities and mobility flows. Second, we present and discuss the results and their implications. Third, we summarize the key insights from our analysis and discuss potential avenues for future work.

\section*{Materials and methods}

With the objective of characterizing the mobility of the residents in a city, here we introduce the datasets and methods used in this paper. Then, we describe how to estimate accessibility to several types of amenities and services within each neighborhood, and how these measurements, jointly with census and mobile phone data, enable the identification of local patterns. Finally, as an example application, we use these patterns to explore the relationship between mobility and rental prices.

\subsection*{City data}

\begin{figure}[!h]
\includegraphics[width=\linewidth]{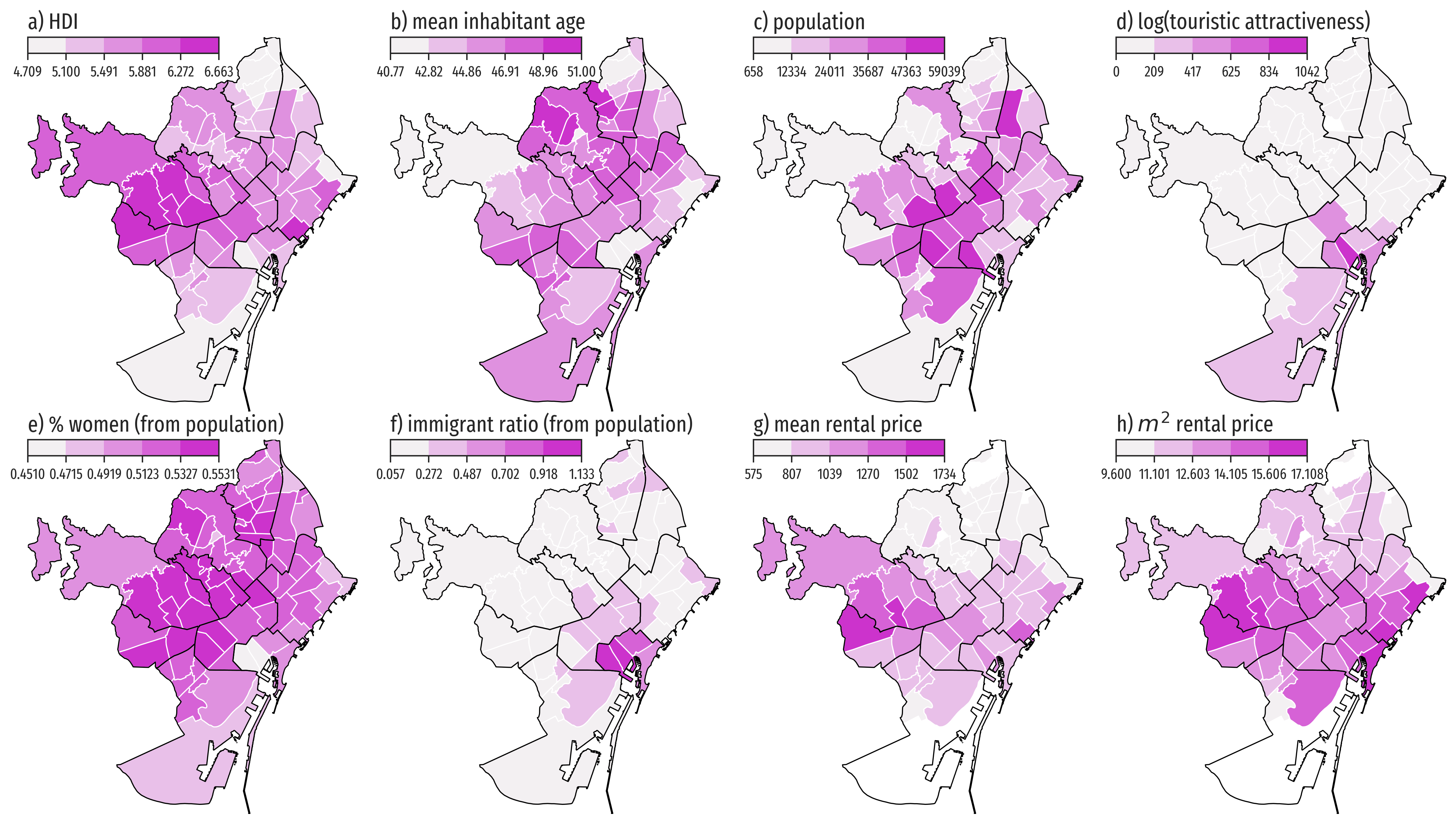}
\caption{{\bf City data for the neighborhoods of Barcelona.}
Choropleth maps of neighborhood features in Barcelona. 
a) Mean socio-economic characterization, Human Development Index (HDI; unitless, from 0 to 10). b) Mean age (years). c) Resident population count. d) Touristic attractiveness. e) Population percentage registered as women. f) Ratio between immigrants and nationals. g) Mean rental price in Euros. h) Mean rental price per square meter in Euros.
Administrative boundaries are sourced from CartoBCN under a CC BY 4.0 license, with permission from Ajuntament de Barcelona, original copyright 2020.
}
\label{fig:bcn_neighborhoods}
\end{figure}

The Barcelona municipality, part of the Barcelona Metropolitan Area, is divided into 73 neighborhoods.
The city data consist of the census available from the \emph{Ajuntament de Barcelona} (city council) and the \emph{Instituto Nacional de Estadísticas} (INE, National Statistics Institute). For each neighborhood we have the following mean values for 2018: a number that characterizes income and social well-being denoted Human Development Index (HDI), age, population, share of population registered as women, share of immigrant population, mean rental price, and mean rental price per square meter (see Fig~\ref{fig:bcn_neighborhoods} for their spatial distributions). 
The distributions of each variable reveal large spatial heterogeneity in income, rental price, touristic attractiveness (defined as log-transformed visitation density from mobile phone data during August 2018, see next subsection), and immigration ratio. These variations suggest that a local approach to understand mobility may be suitable. 

\subsection*{Aggregated mobile phone data flows}

The geographically aggregated mobile phone data were provided by the Barcelona city council, originally collected by Vodafone, one of the largest mobile phone operators in Spain (estimated 25\% to 30\% market share). The dataset contains the aggregated number of visitors with registered active mobile phone operations within the city limits in four-hour time frames, daily. 
Visitors were segregated by the operator in the following types: residents per neighborhood in the Barcelona municipality, commuters from nearby cities, regional tourists, national tourists, and foreign tourists. 
Our work focuses on residents, for whom we can map the patterns in destination visits according to their neighborhood of origin.

We focus our analysis on the {\em most neutral} month (i.e., neither winter or summer breaks, nor containing too many holidays or special events). The selected month is June for two reasons. First, it is a neutral month with no special events (see S1 Appendix). Second, discarding special months, it presents the best goodness-of-fit for the mobility models (details on the next section).

In geographical terms, our dataset contains two types of geolocation: while most of the city data (previous subsection) is associated to neighborhoods or political boundaries, the visitor counts are aggregated in a regular grid (212 cells of $500~m^2$, see details in Ref.~\cite{graells2020measuring}), thus defining a bipartite origin-destination (OD) graph between neighborhoods of residential origin and cells of visitor destination. To work with a neighborhood-to-neighborhood network, we assigned the number of visitors in grid cells to their corresponding neighborhoods by weighting the total number of visitors by the intersection of each cell with the neighborhoods. Note that the weight is calculated by the area of the intersection, not by its population, as the mobility data contains visitor counts to each cell. This count is not necessarily related to the static population of each neighborhood or cell. 

We are interested in the average mobility within a month, as we seek to find generalizable insights. To do so, we modelled the average daily origin-to-destination mobility count. 
There is a remarkable number of null OD pairs (61\% on average), which may be explained due to anonymization purposes in the data: if a cell contained less than 50 people in a given time-frame, it was reported as 0. This implies that some infrequent flows have zeroes in both observations for business days and weekends. We removed those OD pairs, and only kept zeroes when either business days/weekends was non null, as this may characterize work-centric areas, among other situations. The filtered amount of zeroes was reduced to 53\% on average (see the obtained distribution of OD counts in Fig~S2 in S1 Appendix). 

\begin{figure}[!h]
\begin{adjustwidth}{-2.25in}{0in}
\includegraphics[width=\linewidth]{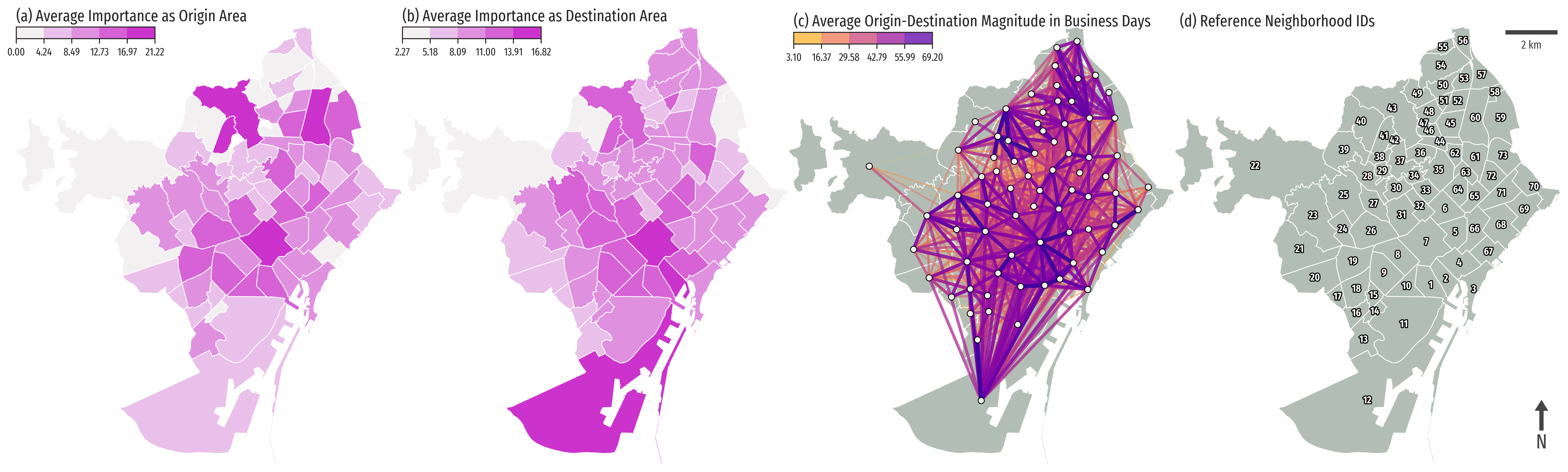}
\caption{{\bf Origin-destination flows between neighborhoods in Barcelona.}
a) Map of average importance as origin area (neighborhoods). b) Map of average importance as destination area (neighborhoods) for local people in Barcelona. c) Node-link diagram of the flow network, with links color coded by the flow importance (log-transformed, monthly average in business days). d) The official reference neighborhood identifiers of the city. 
Administrative boundaries are sourced from CartoBCN under a CC BY 4.0 license, with permission from Ajuntament de Barcelona, original copyright 2020.}
\label{fig:neighborhoods_and_grids}
\end{adjustwidth}
\end{figure}

For illustration purposes, we estimated an overall importance of origin and destination for each neighborhood, defined as the square root of the mean of the number of  incoming/outcoming flows in business days (see Fig~\ref{fig:neighborhoods_and_grids} (a) and (b)). 
The overall destination popularity highlights the historical center, but also the port area, which has almost no inhabitants, but receives many visits due to workers, and people going to/from the airport. We also estimated a popularity score for August 2018 (summer holidays) considering only roaming-plan visitors, and considered it as a touristic attractiveness score for each neighborhood due to being the month with the highest tourist influx in the city (see its spatial distribution in Fig~\ref{fig:bcn_neighborhoods} (d)).
The resulting network has a few noticeable details, such as the predominance of flows between contiguous neighborhoods, and two potential communities that split the city into a north cluster, and a cluster that comprises the downtown and south areas, including the low populated sectors around the port, and the highway connecting with the airport (see Fig~\ref{fig:neighborhoods_and_grids} (c)).

\subsection*{Accessibility in 15 minutes}

Accessibility can be interpreted in different ways depending on the context. Commonly, it is an assessment of the availability of services for older, disabled or disadvantaged people~\cite{Schmocker2008}; as well as reachability to certain places, workplaces or leisure activities during certain times of the day via specific types of transportation~\cite{RoyalSociety}. %
It can be measured in a variety of ways~\cite{RoyalSociety}, from multiple points of reference that include the individual perspective, the place perspective, or both by appropriate population segmentation strategies~\cite{Weng19}. 
Here we focus on the walking accessibility, which has gained increased attention since Carlos Moreno kick-started the 15-minutes city concept in 2019~\cite{Moreno}, and is fueled~\cite{c40} by the potential advantages for urban health~\cite{Alfonzo}, sustainability~\cite{Capasso}, and overall quality of life~\cite{nuvolati}. 
Note that there are alternatives to this proposal. For instance, given that other modes of transportation (such as bicycles or public transport) afford varying levels of accessibility~\cite{Boisjoly,texasassesment}, we could integrate them in the definition. Another option would be to tune the somewhat arbitrary 15-minutes threshold (e.g., Portland pursues accessibility within 20 minutes~\cite{portland}), or weighting amenity types against demographic groups~\cite{Weng19}. We prefer our choice of metric for being clear and straightforward, which will help us understand and communicate how our methods work. Thus, we define the accessibility $A$ to an amenity category $c$ at neighborhood $i$ as follows:
$$
A_{ic} = \frac{\sum_{x \in i} \text{number of } c \text{ amenities reachable by } x \text{ within a 15-minutes walk}}{\text{Population of neighborhood } i},
$$
where $x \in i$ denotes an inhabitant of neighborhood $i$.
The census data contains an approximate address for each inhabitant, which enables the measurement of how many amenities are reachable within a 15-minutes walk.

To quantify the accessibility of each neighborhood, we downloaded an OSM~\cite{OpenStreetMap} database dump of 2018 to identify POIs in the city, as well as the walkable street network of the city. OSM is a Volunteered Geographical Information site where anyone can enter and edit geographical information. It has a similar (and sometimes greater and up-to-date) coverage and accuracy to commercial maps~\cite{haklay2010good}, and it is freely available.
We calculated the accessibility metric for the following amenity categories: \emph{education} (schools, universities, etc.), \emph{entertainment} (indoor places where people pay for access), \emph{finance} (banks and similar), \emph{food} (restaurants, bars, cafes, etc.), \emph{government} facilities, \emph{health} facilities (hospitals, primary care centers, medical offices, etc.), \emph{professional} services, \emph{recreational} areas (parks, outdoors, etc.), \emph{religion} venues, \emph{retail} (malls, supermarkets, shopping venues), and \emph{public transport} (bus stops, metro stations, bike stations, etc.). 
We also estimated the diversity of reachable amenities per inhabitant, as diversity has shown to be an important predictor in characteristics and development of places~\cite{eagle2010network,quercia2018diversity}. The diversity is defined as the Shannon Entropy:
$$
H = - \sum_{c \in C} p_{c} \log p_{c},
$$
where $p_c$ is the frequency of an amenity of category $c \in C$, and $C$ is the set of amenity categories, excluding \emph{public transport}. 

\begin{figure}[!h]
\includegraphics[width=\linewidth]{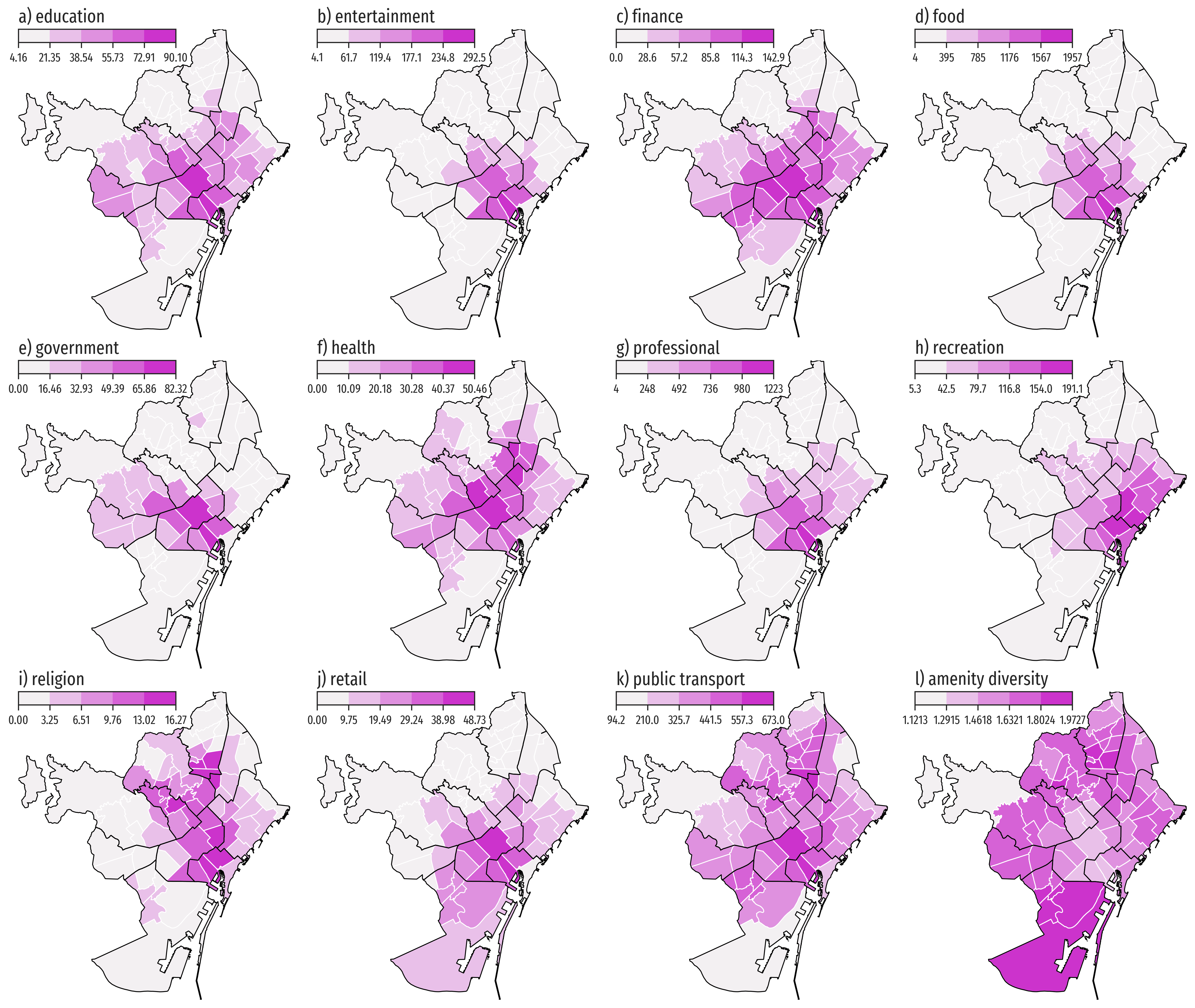}
\caption{{\bf Distribution of mean number of reachable amenities (and its diversity) per neighborhood.}
Each map shows the spatial distribution of the accessibility to the following type of amenities: a) education. b) entertainment. c) finance. d) food. e) government. f) health. g) professional. h) recreation. i) religion. j) retail. k) public transport. The accessibility value per neighborhood corresponds to the mean of the accessibility values of the neighborhood inhabitants. Panel l) shows the diversity of reachable amenities in each neighborhood.
Administrative boundaries are sourced from CartoBCN under a CC BY 4.0 license, with permission from Ajuntament de Barcelona, original copyright 2020.}
\label{fig:amenity_distribution}
\end{figure}

The spatial distribution of these metrics showcase how the central neighborhoods concentrate a greater amount of amenities, which could indicate a monocentric city. However, the diversity of amenities follows an opposite trend, hinting that the city provides (or aims to provide) access to several types of venues and services in its entire administrative area (see their spatial distributions in Fig~\ref{fig:amenity_distribution}).

For each neighborhood in the city, we now have several metrics related to its population, how this population moves in the city from origins to destinations, and the amenities that are reachable from a person's place of usual residence. We now set out to estimate global mobility patterns in the city.

\subsection*{Global mobility patterns from a gravity model}

At city scale, the most common strategy to model flows is to fit a form of gravity model~\cite{zipf1946p}. 
Gravity models are a simple phenomenological approach that allows us to quantify the movement between two zones of different sizes taking into account only three parameters: the population of the origin and destination regions, and the distance between them.
The reasoning behind this is that mobility between two regions should be proportional to their population sizes, but inversely proportional to the distance between the regions, to account for the inconvenience of moving between distant locations.

Gravity models are not the only way to model mobility. One of the most recent models is the radiation model~\cite{simini2012universal}, which is parameter free and provides better prediction performance for commuting flows between cities and counties. In contrast to the gravity model, originally proposed as a metaphor, the radiation model has coherent units and can be derived from human behavior. However, despite these positive traits, the gravity model is still preferred by urban modelers and practitioners. Being parameter free, the radiation model makes it difficult to establish the relationship between mobility flows and place-specific features. Moreover, the gravity model is flexible to incorporate additional explanatory variables. This flexibility may explain its comparatively better performance in inferring intra-city flows~\cite{liang2013unraveling,palchykov2014inferring}. As such, understanding why the gravity model works, and how it relates to human behavior and other theory-based models is an active area of research~\cite{hong2019gravity}. 
For these reasons, we work with gravity models as it offers improved flexibility and performance in our context.

A gravity model is typically defined as follows:
$$
T_{ij} = \frac{G M_i^\alpha M_j^\beta}{f(d_{ij})^\gamma},
$$
where $T_{ij}$ is the number of visitors at destination area $j$ from origin area $i$, $G$ is a gravity constant, $M_i$ is a measure of mass for area $i$ (usually its population), and $f$ is a deterrence function based on the distance $d_{ij}$ between areas $i$ and $j$. A common and well-behaved way to find the parameters of this model is by adjusting a Generalized Linear Model (GLM), with a Negative Binomial (NB) distribution for count data. This distribution has been used before to model area influx in mobile phone data~\cite{graells2017effect,beiro2018shopping}, as it controls for over-dispersion. The model is specified as follows:
$$
\mathbb{E}[T_{ij}] = \exp[\log(G) + \alpha \log(M_i) + \beta \log(M_j) - \gamma \log(f(d_{ij}))].
$$
This GLM is fitted by maximizing the log-likelihood function.

To account for the socio-demographic and accessibility attributes of each neighborhood, we generalize the model equation into:
$$
\mathbb{E}[T_{ij}] = \exp[\beta_0 + \sum_k \beta_{kij} x_{kij}],
$$
where $x_{kij}$ is the $k$th-feature of the flow from neighborhood $i$ to neighborhood $j$. These features include:

\begin{itemize}
    \item Population of origin and destination areas, and the distance that separates them.
    \item Touristic attractiveness for origin and destination areas.
    \item The difference in accessibility between origin $i$ and destination $j$: $\Delta A_{ijc} = A_{jc} - A_{ic}$, where $c$ is an amenity category, $A_{jc}$ is the accessibility to $c$ venues at neighborhood $j$ (the definition of $A_{ic}$ is analogous). A positive value indicates that the people from neighborhood $i$ moves toward a neighborhood $j$ with better accessibility in the corresponding category.
    \item Similarly, we estimate the difference in amenity diversity, $\Delta \text{amenity diversity}_{ij}$.
    \item The difference in socio-demographic characterization $\Delta \text{HDI}_{ij}$ between two areas.
    \item The difference in the fraction of the population that is women ($\Delta \%\text{women}_{ij}$).
    \item The difference in the ratio of population between immigrants and nationals ($\Delta \text{immigrant}_{ij}$).
    \item The difference in mean age of the destination and origin areas, $\Delta \text{age}_{ij}$,
    \item A binary variable $\text{weekend}$ indicating if the visitor count from $i$ to $j$ corresponds to a weekend day (1) or not (0).
    \item A binary variable $\text{to other neighborhood}$ indicating if $i \neq j$ in the corresponding $T_{ij}$.
\end{itemize}

Before fitting the model, we log-transform the distance and population variables (including touristic attractiveness). 
The results of fitting the regression model using these features allows us to measure how the different factors are related to the mobility in the city, as it is usually done in related works. 

\subsection*{Local mobility patterns using Geographically Weighted Regression (GWR)}

If our premise --that a city is a system of many internal cities-- is correct, we should observe that a global model misses peculiarities for each sub-system of the city. We would expect that these cities were different due to their local cultures, location within the city, and other factors. For instance, distance as a factor may not be equally relevant for everyone, as some people may need to travel longer distances to work. Sometimes the factor may have opposite effects, i.e., a factor that is associated to out-neighborhood mobility in a neighborhood is associated to in-neighborhood mobility in another; for example, accessibility to entertainment amenities might attract people from all over the city to a particular area. All these situations might not be well captured in a global model, because it would try to enforce a constant factor for the entire city.

To account for local variations that affect mobility within the city, the model should include a notion of geographical position in its parameters.
One effective approach to this are Geographically Weighted Regression (GWR) models~\cite{brunsdon1996geographically}. GWR is a kernel regression approach to geographically varying phenomena that performs several GLM regressions~\cite{Fotheringham2002}, one per spatial unit under analysis. Each of these regressions weights the observations in the data (in our case, origin-destination flows) according to the distance to the spatial unit in question, through a kernel function that estimates the weight according to a given bandwidth. The bandwidth can be defined either based on domain knowledge or using model-selection criteria~\cite{comber2020}.

A GWR model is expressed as follows:
$$
\mathbb{E}[T_{ij}] = \exp[\beta_{0ij} + \sum_k \beta_{(k)}(\mathbf{u}_{ij}) x_{kij}],
$$
where $\beta_{(k)}(\mathbf{u}_{ij})$ is a vector of geographically varying parameters at a given position $\mathbf{u}_{ij}$ of $T_{ij}$, and $\beta_{0ij}$ is the intercept corresponding to $T_{ij}$. Since we work with origin-destination flows, there are three alternatives to define $\mathbf{u}_{ij}$: the origin of each OD pair, its destination, or its midpoint~\cite{blainey2013extending}. We chose to consider the origin position given our problem definition: we are interested in patterns at the origin neighborhoods, i.e., what motivates people to move from or to stay at a given area. Patterns in destination neighborhoods may be of use to estimate land use patterns, which is out of the scope of this paper, or the mirror question of what attracts people to a given area. OD mid-points make little sense in this context as two orthogonal flows may share their midpoint without having anything else in common. In contrast, the origin position allows agglomeration of all OD pairs starting at our unit of analysis.

The outcome of applying GWR is a vector of regression factors for every OD pair, which can be aggregated by neighborhood. By definition, all OD pairs with the same origin position will have the same regression results because they use the same spatial kernel. We interpret the resulting vectors as a characterization of each neighborhood's  mobility and accessibility. 
If we observe little spatial variation in the coefficient vectors this would refute our city of cities assumption. On the other hand, significant spatial variation of the coefficient vectors would indeed support the idea of multiple cities within a city, each one with their own mobility patterns. 

\subsection*{Relationship between local mobility and rental prices}

To exemplify how local patterns of mobility can improve our understanding of cities, we explore how these patterns relate to an existing problem in Barcelona, namely, the surge in rental prices. The cost of life in a city is deeply attached to this value, and Barcelona has recently faced over-tourism~\cite{milano2019overtourism} and gentrification processes~\cite{anguelovski2018assessing}.
The problem is well researched, in terms of which local factors of neighborhoods are related to sudden surges in price and tourist accommodation, and its consequences on housing prices~\cite{blanco2018barcelona,garcia2019short,lagonigro2020understanding}. 

Typical pricing modeling uses hedonic regression~\cite{halvorsen1981choice}. In addition to housing unit attributes, neighborhood and population features are also included in such models, as well as the spatial dependency of the issue under analysis.
Up to now, the phenomena has been explored from a static perspective where neighborhoods only have contiguity relationships and land use designations. However, how people connect neighborhoods through their mobility is also important. With relationships like these, urban planners and the city council would have more information to consider in their plans to improve quality of life in the city.

Here we propose to measure this relationship using correlation. The Spearman correlation coefficient $\rho$ between two vectors is defined as:
\begin{equation*}
    \rho(X, Y) = \frac{\text{cov}(X_R, Y_R)}{\sigma_{X_R} \sigma_{Y_R}},
\end{equation*}
where \textit{cov} is the covariance between two vectors, and $\sigma$ is the standard deviation of a vector. In this case, the functions are applied to the rank-transformed versions of $X$ and $Y$, denoted $X_R$ and $Y_R$. The value of $\rho$ lies in $[-1, 1]$, and it is interpreted as follows: a value of $\rho = 0$ shows no relation between the two variables; a value of $\rho = 1$ shows a perfect positive relation (i.e., an increase in one variable is reflected with an equivalent increase in the other); and a value of $\rho = -1$ shows a perfect negative relation where increases in one variable are related to decrements in the other. Thus, we calculate the $\rho$ coefficients between the rental attributes of each neighborhood and its local mobility variables, and analyze both their magnitude and sign to find which local patterns are related with higher or lower rental princes in the city.

\section*{Results}

Being a vibrant city, Barcelona has several periods throughout the year that influence how people move within it. For instance, the Mobile World Congress (represented by the highest spike in February in Fig~S1 in S1 Appendix) brings many foreigners and national visitors into the city, and changes to the transportation network are common to accommodate the increased demand on transportation. Another example is that an important fraction of the resident population leaves the city during holiday periods, especially in August (see also Fig~S1 in S1 Appendix). 
This seasonality suggests that we need to define a typical or average month to analyze mobility. Hence, for every month of the year, we estimated the average origin-destination flow between neighborhoods at business days and weekends, and fitted a global model on them. Our criteria to select a month to analyze was to compare the Akaike Information Criterion corrected (AICc) of each model. The AICc is a measurement of information loss, as such, a model with lower AICc is preferred. The corrected version of AIC accounts for multiple model comparisons and their sample size. In this setting, the months with the lowest AICc are August, December and June, in that order (see Fig~S3 in S1 Appendix). However, considering seasonality, the two months with lowest AICc values (August and December) exhibit a decay of local population due to the summer holidays and the Christmas break, and thus, do not represent a typical month of city life. Hence, we chose the third lowest value: June. This month will be used for the rest of the analysis.

\begin{figure}[!h]
\includegraphics[width=\linewidth]{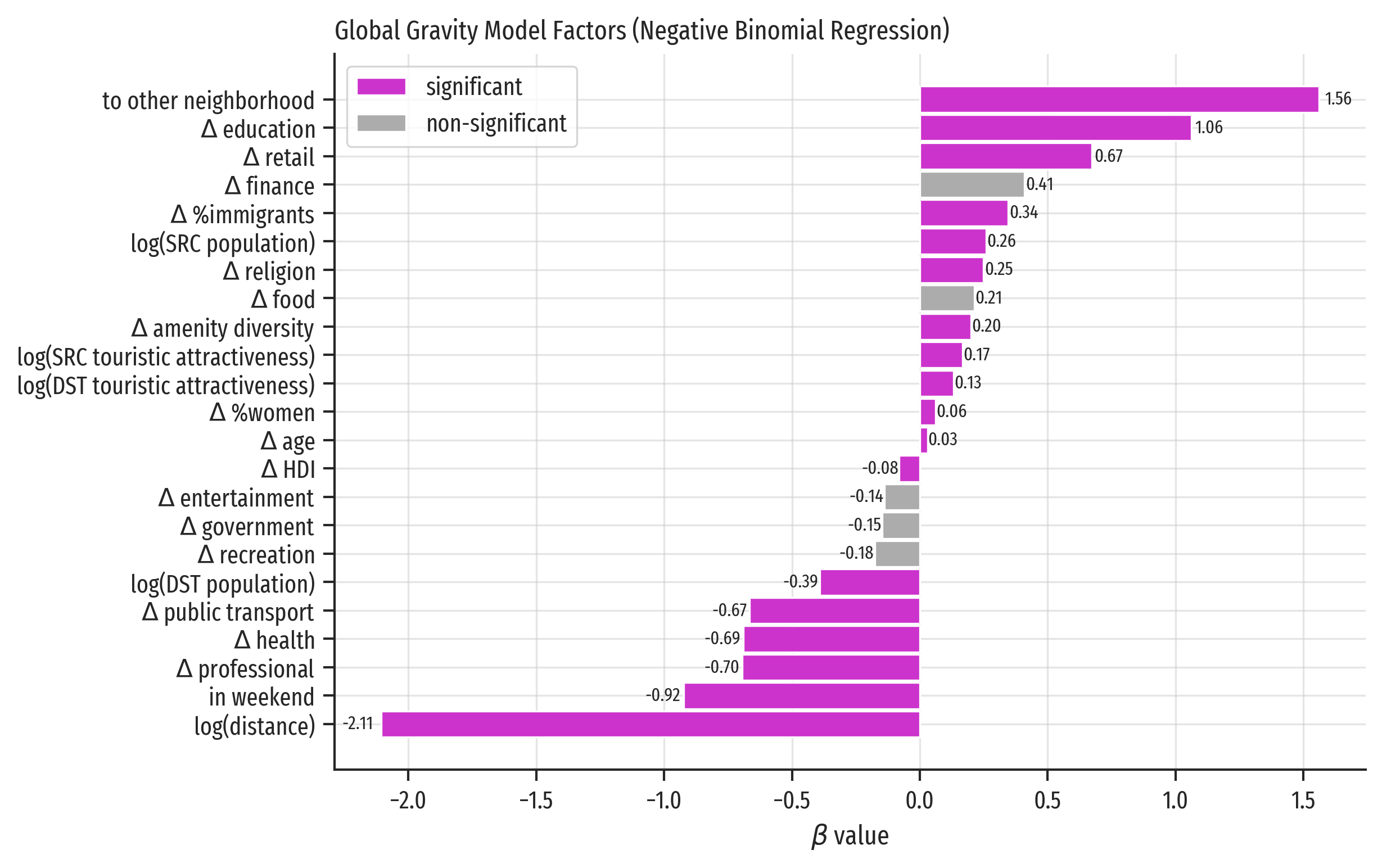}
\caption{{\bf Global gravity model factors.}
Positive values indicate that the corresponding factor is associated with an increase in flows from one neighborhood to another.}
\label{fig:global_model_results}
\end{figure}

In the global model the two most important factors are: \textit{log(distance)} ($\beta = -2.11$), with a negative effect, i.e., longer distances between neighborhoods imply smaller flows; and \textit{to other neighborhood} ($\beta = 1.56$), with a positive effect, implying that flows from a neighborhood tend to go to other neighborhoods (see all factors in Fig~\ref{fig:global_model_results}). Attributes such as $\Delta$~\textit{education}, $\Delta$~\textit{retail}, and $\Delta$~\textit{finance} are positive. To interpret these values, recall that $\Delta$ variables were defined as the difference between a destination value minus an origin value. Thus, a positive factor can be interpreted as follows: at the city level, people tend to visit neighborhoods with greater access to \textit{education} and \textit{retail} than their own. On the contrary, there is a tendency to travel less to neighborhoods when they have greater accessibility to \textit{professional} and \textit{health} amenities. The former type of venue offers services provided by professionals, and may not be remarkably different between neighborhoods. The health amenities focus on local people by design of the public health system, CatSalut. These insights are expected and coherent with the literature and the urban design of Barcelona. 

Before adjusting the local model, we needed to select a kernel function and a bandwidth for the ratio of influence between neighborhoods and their flows. 
The tested kernels were exponential and bi-square distance weight decay, and the bandwidth ranged between 750 meters and 2 kilometers (with specific values tested using bisection search). We kept the parameters from the model with the lowest AICc~\cite{oshan2019mgwr}. The final local model was fitted using 750 meters as bandwidth and an adaptive bisquare kernel.
To check model coherence, we noted that the model residuals present a symmetric distribution around zero in both cases, global and local (see Fig~S4 in S1 Appendix). 

\begin{figure}[!h]
\begin{adjustwidth}{-2.25in}{0in}
\includegraphics[width=0.95\linewidth]{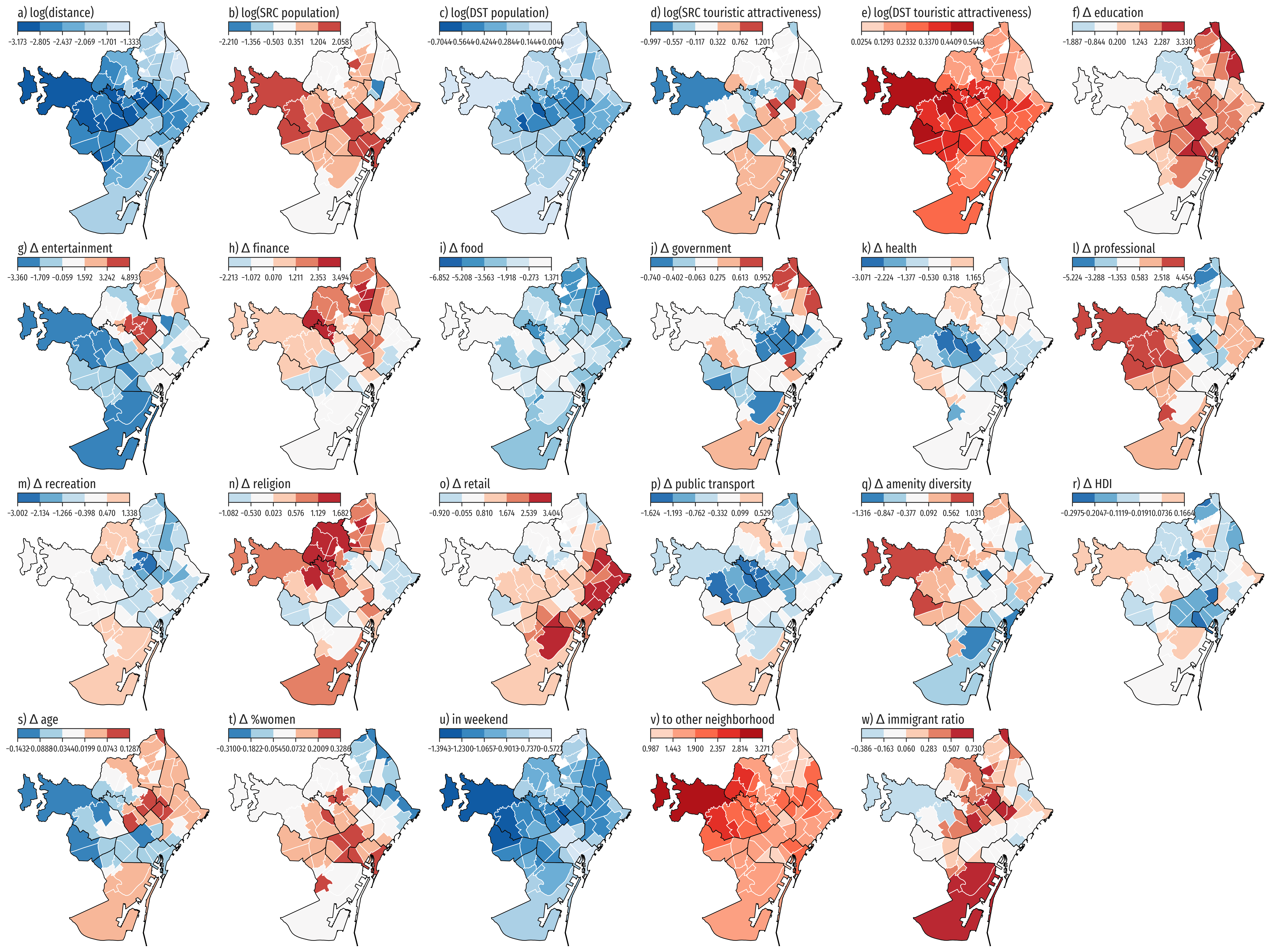}
\caption{{\bf Spatial distribution of the local regression factors from the GWR model.}
Each choropleth map uses a color scale with red tones to depict positive values, and blue tones to depict negative values. Factors: a) Distance. b) Neighborhood Population. c) Destination Population. d) Neighborhood Touristic Attractiveness. e) Destination Touristic Attractiveness. f) $\Delta$~\emph{education}. g) $\Delta$~\emph{entertainment}. h) $\Delta$~\emph{finance}. i) $\Delta$~\emph{food}. j) $\Delta$~\emph{government}. k) $\Delta$~\emph{health}. l) $\Delta$~\emph{professional}. m) $\Delta$~\emph{recreation}. n) $\Delta$~\emph{religion}. o) $\Delta$~\emph{retail}. p) $\Delta$~\emph{public transport}. q) $\Delta$~\emph{neighborhood amenity diversity}. r) $\Delta$~\emph{neighborhood Human Development Index} (HDI). s) $\Delta$~\emph{neighborhood mean age}. t) $\Delta$~\emph{\% women population}. u) weekend flag (binary). v) visiting other neighborhood (binary). w) $\Delta$~\emph{ratio immigrants to nationals}.
Administrative boundaries are sourced from CartoBCN under a CC BY 4.0 license, with permission from Ajuntament de Barcelona, original copyright 2020.}
\label{fig:gwr_model_results}
\end{adjustwidth}
\end{figure}

We report global and local model statistics and their factor values in Table~\ref{table:model_results}. 
The local model AICc is considerably lower than that for the global model, suggesting that our local model can better explain human mobility flows in Barcelona. The GWR results reveal wide geographic variability in the relationship between human mobility flows and local factors. GWR regression estimates range from negative to positive and display relatively high standard deviations. Only five variables are consistently negative ($\log$(\textit{distance}), \textit{in weekend} and $\log$(\textit{destination population})), or positive (\textit{to other neighborhood} and $\log$(\textit{destination touristic attractiveness})). These estimates show the expected sign and are consistent with our global model estimates. However, the extent of the relationship between these factors and human mobility flows varies across Barcelona (Fig~\ref{fig:gwr_model_results}). Distance, for example, seems to exert a prominently large impact on deterring mobility in western and central-western neighborhoods but has a less pronounced influence on coastal northern and southern neighborhoods.

\begin{table}[!ht]
\begin{adjustwidth}{-2.25in}{0in} % Comment out/remove adjustwidth environment if table fits in text column.
\centering
\caption{
{\bf Regression Results in the Global Model (Negative Binomial Regression) and the Local Model (GWR with GLM Negative Binomial Regression).}}
\begin{tabular}{|p{4.1cm}|rrrr|rrrrr|}
\hline
{} & \multicolumn{4}{c|}{\textbf{Global Model} (AICc = 10544.52)} & \multicolumn{5}{c|}{\textbf{Local Model (GWR)} (AICc = 5625.56)}\\ 
                                 Variable &    $\beta$ &     SE &  t(Est/SE) &      $p$-value &    Mean &     STD &      Min &  Median &     Max \\
\thickhline
                                Intercept &  3.638 & 0.300 &     12.128 & $<$ 0.001 & -2.059 & 7.122 & -19.758 &  -0.857 & 22.533 \\
                   $\Delta$ education &  1.061 & 0.167 &      6.340 & $<$ 0.001 &  1.188 & 1.078 &  -1.887 &   1.331 &  3.325 \\
               $\Delta$ entertainment & -0.139 & 0.196 &     -0.707 & 0.479 &  0.106 & 1.741 &  -3.360 &  -0.122 &  4.884 \\
                     $\Delta$ finance &  0.409 & 0.231 &      1.766 & 0.077 &  0.512 & 1.462 &  -2.213 &   0.731 &  3.488 \\
                        $\Delta$ food &  0.213 & 0.245 &      0.870 & 0.384 & -1.907 & 1.710 &  -6.852 &  -1.932 &  1.363 \\
                  $\Delta$ government & -0.147 & 0.084 &     -1.743 & 0.081 & -0.027 & 0.388 &  -0.740 &   0.006 &  0.950 \\
                      $\Delta$ health & -0.691 & 0.186 &     -3.709 & $<$ 0.001 & -0.589 & 0.906 &  -3.071 &  -0.434 &  1.161 \\
                $\Delta$ professional & -0.696 & 0.313 &     -2.219 & 0.026 & -0.137 & 2.083 &  -5.224 &  -0.083 &  4.444 \\
            $\Delta$ public transport & -0.666 & 0.090 &     -7.419 & $<$ 0.001 & -0.456 & 0.507 &  -1.624 &  -0.339 &  0.527 \\
                  $\Delta$ recreation & -0.177 & 0.149 &     -1.190 & 0.234 & -0.467 & 0.883 &  -3.002 &  -0.566 &  1.334 \\
                    $\Delta$ religion &  0.249 & 0.100 &      2.498 & 0.012 &  0.131 & 0.736 &  -1.082 &  -0.012 &  1.679 \\
                      $\Delta$ retail &  0.672 & 0.126 &      5.347 & $<$ 0.001 &  1.296 & 0.913 &  -0.920 &   1.270 &  3.400 \\
                  $\Delta$ amenity diversity &  0.199 & 0.094 &      2.128 & 0.033 & -0.146 & 0.401 &  -1.316 &  -0.062 &  1.029 \\
                        $\Delta$ HDI & -0.082 & 0.030 &     -2.728 & 0.006 & -0.041 & 0.100 &  -0.297 &  -0.036 &  0.166 \\
                        $\Delta$ age &  0.031 & 0.012 &      2.615 & 0.009 &  0.003 & 0.068 &  -0.143 &   0.022 &  0.128 \\
                 $\Delta$ \%women &  0.063 & 0.028 &      2.217 & 0.027 &  0.042 & 0.152 &  -0.310 &   0.054 &  0.328 \\
                  $\Delta$ \%immigrants &  0.345 & 0.056 &      6.194 & $<$ 0.001 &  0.192 & 0.259 &  -0.386 &   0.193 &  0.729 \\
                     log(distance) & -2.106 & 0.041 &    -51.802 & $<$ 0.001 & -2.328 & 0.452 &  -3.173 &  -2.312 & -1.335 \\
         log(DST population) & -0.392 & 0.027 &    -14.390 & $<$ 0.001 & -0.300 & 0.145 &  -0.704 &  -0.293 & -0.005 \\
 log(DST touristic attractiveness) &  0.132 & 0.021 &      6.449 & $<$ 0.001 &  0.279 & 0.110 &   0.025 &   0.320 &  0.544 \\
         log(SRC population) &  0.259 & 0.031 &      8.229 & $<$ 0.001 &  0.663 & 0.738 &  -2.210 &   0.617 &  2.054 \\
 log(SRC touristic attractiveness) &  0.166 & 0.020 &      8.230 & $<$ 0.001 &  0.178 & 0.387 &  -0.997 &   0.111 &  1.199 \\
                       to other neighborhood &  1.560 & 0.105 &     14.865 & $<$ 0.001 &  1.781 & 0.364 &   0.987 &   1.794 &  3.269 \\
                                  in weekend & -0.923 & 0.032 &    -28.661 & $<$ 0.001 & -1.034 & 0.169 &  -1.394 &  -1.066 & -0.573 \\
\hline
{Local Model Info.} & \multicolumn{9}{c|}{Adaptive bisquare kernel, Bandwidth=769m, Adj. critical t value (95\%)=2.95, DF=4380.72}\\
\hline
\end{tabular}
\begin{flushleft} Summary table for global and local regression models. Acronyms and abbreviations: AICc (Akaike Information Criterion corrected), SRC (source neighborhood), DST (destination neighborhood), HDI (human development index). In the local model, \textit{Mean} refers to the mean $\beta$ value of all coefficients for all regressions performed within the model. The other statistical properties are defined analogously.
\end{flushleft}
\label{table:model_results}
\end{adjustwidth}
\end{table}

A much wider extent of geographic variability exists in the relationship between human mobility and other factors, and for many neighborhoods, this relationship contrasts with that captured by the global model. For instance, the global model indicates that differences in educational and retail amenities between neighborhoods greatly contribute to shaping mobility. Yet, our GWR estimates indicate that while disparities in accessibility to educational facilities are strongly associated with mobility out of certain central and northern neighborhoods, and with reduced mobility in others, this characteristic is more prominent in southern and north-western neighborhoods. Similarly, GWR estimates indicate that while unequal access to retail amenities tend to promote mobility from central and eastern coastal neighborhoods, they are negatively associated with mobility in western and northern areas of Barcelona. In addition, our results also reveal how geographic differences in the composition of the resident population is associated with mobility. For example, our GWR estimates indicate that large differences in age profile tend to trigger mobility from inland eastern neighborhoods, suggesting that areas with an older population profile tend to experience higher levels of mobility.

\begin{figure}[!h]
\begin{adjustwidth}{-2.25in}{0in}
\includegraphics[width=\linewidth]{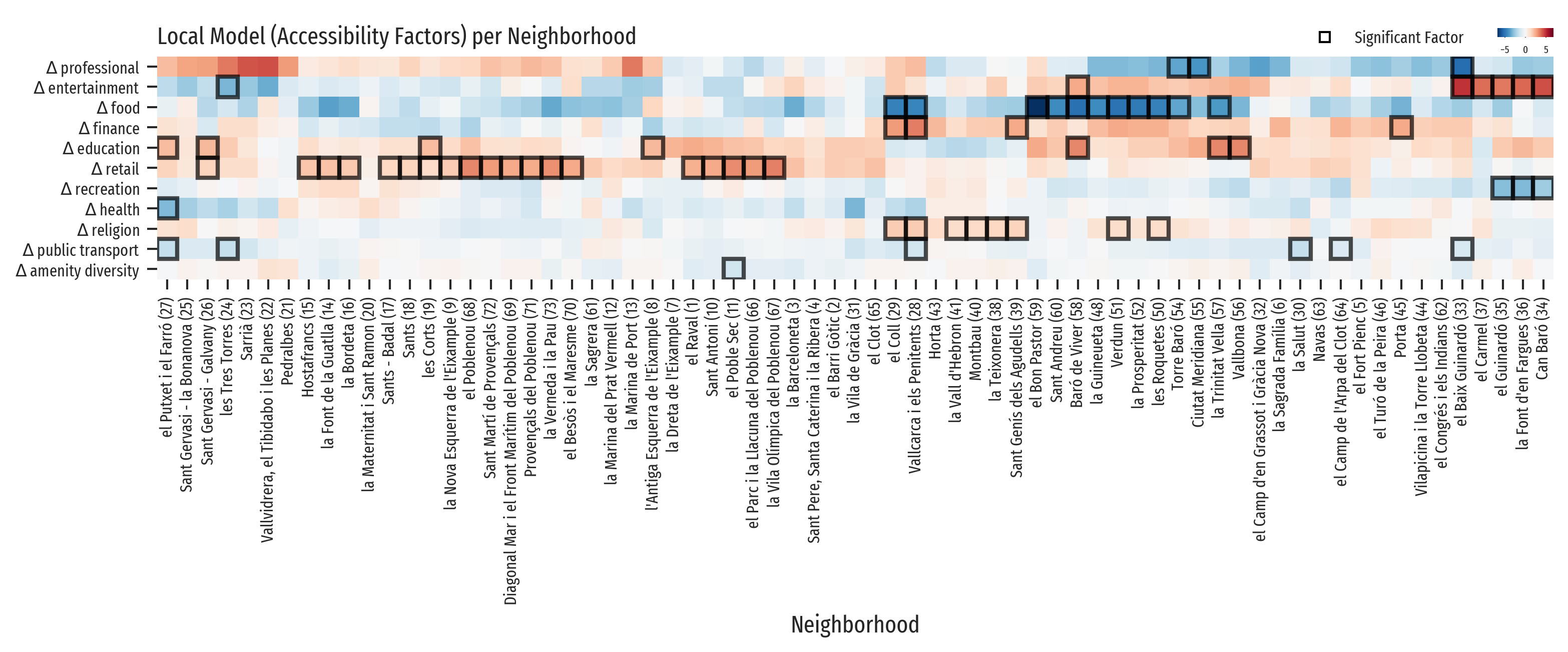}
\caption{{\bf Heatmap of GWR coefficients by neighborhood and variable.} This heatmap shows the results from the GWR model, allowing for the identification of clusters of neighborhoods and variables. The colormap is diverging, negative values are blue, positive values are red, and values close to 0 are white. Variables were sorted by their standard deviation and neighborhoods were ordered and clustered using ward linkage distance. The cells with an overlaid black square are statistically significant at the 0.05 level.}
\label{fig:heatmap_GWR_coefficients}
\end{adjustwidth}
\end{figure}

Polycentric city strategies are associated with highly walkable spaces within local neighborhoods and reduced long-distance mobility to other neighborhoods. Key to inform such strategies is identification of the set of urban amenities influencing mobility from neighborhoods contributing to low local levels of resident population retention. To identify the set of urban amenities underpinning local levels of mobility in Barcelona, we created a heatmap that shows the accessibility factors from the GWR and highlights the significant ones, such that these can provide local actionable insights to experts and policy makers. % xxxx. 
Fig~\ref{fig:heatmap_GWR_coefficients} reveals that a small but distinctive set of urban amenities tend to shape the local patterns of human mobility in most neighborhoods across Barcelona, such as retail, education, and food amenities. The majority of neighborhoods have differences in accessibility to urban amenities, and these tend to significantly influence mobility from different neighborhoods. Differences in accessibility to retail outlets stands out as the most consistent factor linked to increased levels of mobility in many neighborhoods. Interestingly, the reduced levels of mobility effect from differences in food amenity accessibility is probably explained because of the large concentration of food amenities in the center of the city, meaning that the citizens from these neighborhoods do not tend to move to the center of the city. Differences in accessibility to entertainment from neighborhoods from the north-western inner part of the city also increase levels of mobility, similarly to the differences in food amenities, but having a positive effect on the levels of mobility as the data suggests that these people tend to move to the center of the city. Lastly, differences in accessibility to religion amenities from north-western neighborhoods also stand out increasing levels of mobility.
 
\begin{figure}[!h]
\includegraphics[width=\linewidth]{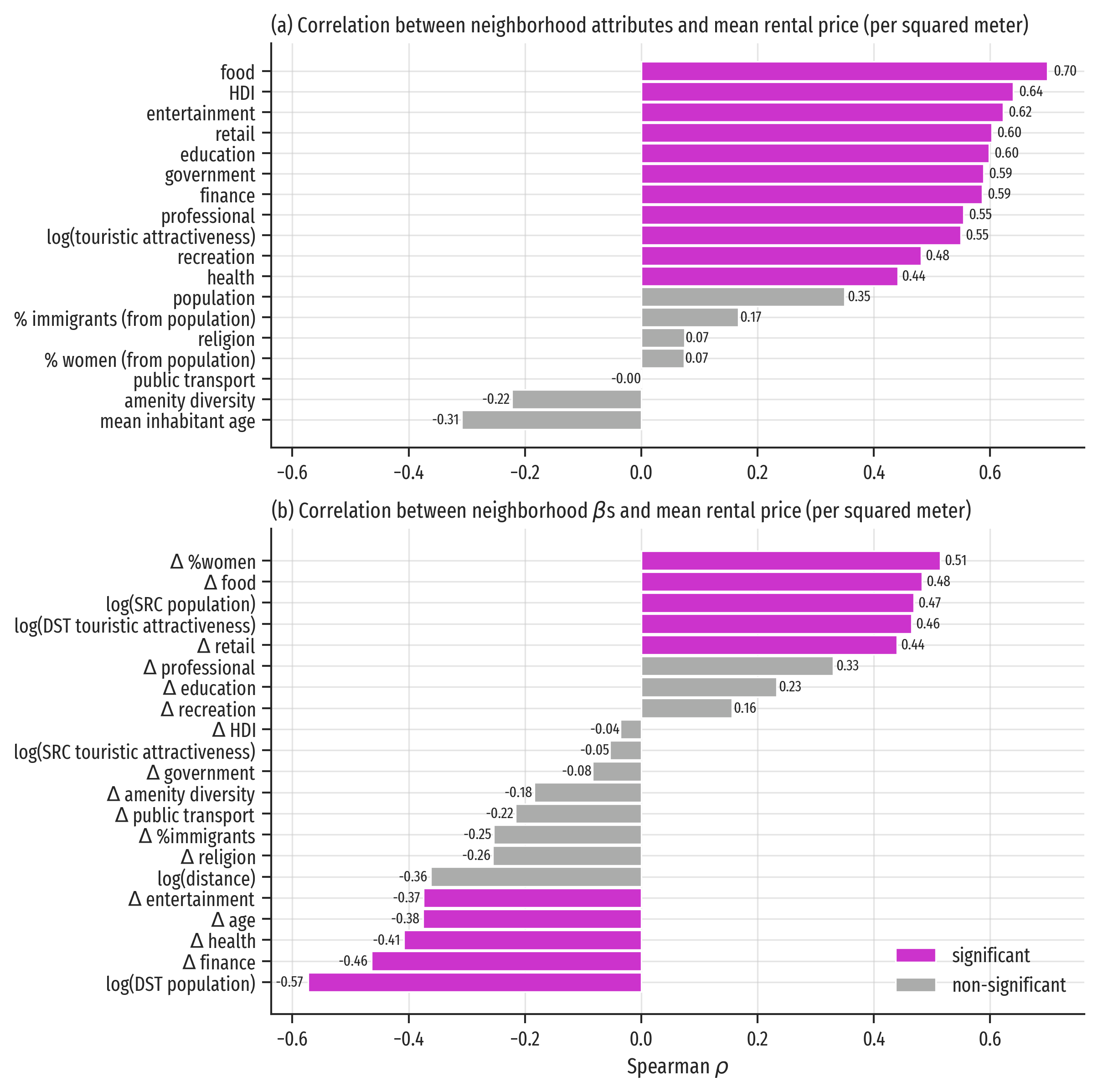}
\caption{{\bf Correlations between neighborhoods' attributes and mobility factors with relative rental prices.} Grey bars represent non-statistically significant correlations, magenta bars represent statistically significant correlations; significance was determined with Bonferroni-corrected $p$-values (significance level 0.05).
a) Spearman correlation between neighborhood features and rental price per square meter. b) Spearman correlation between neighborhood local mobility factors and rental price per square meter.}
\label{fig:mobility_rent_correlations}
\end{figure}

Finally, we explored how the local patterns of mobility allow us to find novel perspectives regarding urban phenomena. We estimated the correlation between relative neighborhood rental prices and the neighborhood characterization, both from its direct features and those derived from the local model of mobility.
Many of the neighborhood attribute correlations with rental prices are significant (see Fig~\ref{fig:mobility_rent_correlations} (a)). The exceptions are access to \emph{religion}, to \emph{public transport}, the amenity diversity, and the demographic variables related to population size, age, gender and migration. The significant correlations are positive (ranging from access to health venues with $\rho = 0.44$ to access to food venues with $\rho = 0.70$), which is expected, as better accessibility in a neighborhood is a positive trait. 

There are ten variables from the local features that have significant correlations with rental prices, five positive and five negative (see Fig~\ref{fig:mobility_rent_correlations} (b)).
The positive correlations are: 
$\Delta$~\emph{\% women} ($\rho = 0.51$),
$\Delta$~\emph{food} ($\rho = 0.48$), 
$\log$(\textit{origin population}) ($\rho = 0.47$),
$\log$(\textit{destination touristic attractiveness}) ($\rho = 0.46$), 
and $\Delta$~\emph{retail} ($\rho = 0.44$). 
Of these variables, three are related to tourism (access to food and retail venues, and the attractiveness of the destination area), which showcases that neighborhoods where its inhabitants go to shop and dine are more expensive. These four variables can be linked to gentrification processes, suggesting that places that are not necessarily touristic but allow to reach touristic places are more expensive for housing, in coherence with over-tourism in the city~\cite{garcia2019short}.
The positive association with population size needs more exploration, as it needs to take into account the different types of residential density.
Regarding the strong positive correlation with the $\Delta$~\emph{\% women} factor, note that according to the spatial distribution of this factor (see Fig~\ref{fig:gwr_model_results}~(t)), there are sign differences between the city core (positive) and its periphery (negative), which is different from the neighborhood attribute of fraction of population that is women (see Fig~\ref{fig:bcn_neighborhoods}~(e)). We interpret this difference in relationship to access to safe places by women, and other potential gender gaps in urban mobility~\cite{gauvin2019gender}.  
The negative correlations include the following factors: $\log$(\textit{destination population}) ($\rho = -0.57$), which means that neighborhoods where its inhabitants tend to travel to densely populated areas have lower rental prices; $\Delta$~\emph{finance} ($\rho = -0.46$) and $\Delta$~\emph{health} ($\rho = -0.41$), showing that areas where inhabitants move to other neighborhoods for these amenities are less expensive with respect to housing. Lastly, the $\Delta$~\emph{entertainment} correlation ($\rho = -0.37$) shows that neighborhoods where its residents move to other neighborhoods for amenities linked to nightlife and tourism in Barcelona have lower rental prices.

In summary, these results provide evidence on how Barcelona is not a monocentric city, but rather a polycentric city with diverse functional areas. Distinct groups of factors influence different regions and their local mobility, and it is the combination of amenities with the different ``small-cities'' within the city that define the many centers of Barcelona. Note that the actual shape of these multiple cities is not defined; such a task is out of the scope of this paper, however, a potential way to achieve this segmentation is through these local representations. As we observed, the local variability in mobility factors may inform urban planners and policy makers regarding the functioning of the city and its on-going phenomena. Besides, even though some variables are not necessarily significant with respect to origin-destination flow inference, we find these are relevant as they present variability and are informative with respect to other issues. 

\section*{Discussion and Conclusions}
We sought to improve our understanding of the relationship between human mobility and urban amenities in the context of polycentric and 15min city planning strategies. Particularly we aimed to measure the extent of geographic variability in this relationship and how it was associated with the socio-demographic profile of local population. We integrated new (mobile phone and OSM data) and traditional (census data) forms of data to model and quantify the relationship between human mobility flows, and urban amenities and local population composition.
 
Our results provided evidence of wide geographic variability in the relationship between human mobility and accessibility to local urban amenities. We showed that a distinctive set of local urban amenities contribute to shape local patterns of human mobility at the neighborhood level in Barcelona. Our findings revealed that the local patterns of human mobility in most neighborhoods in Barcelona are explained by differences in accessibility in one or two types of urban amenities. Unequal access to food outlets emerged as the most consistent factor linked to reduced levels of mobility in most neighborhoods.

We also presented evidence of significant differences between global and local patterns. We showed that accessibility indicators of urban amenities displaying statistically significant relationships with human mobility at the global level do not necessarily explain human mobility patterns at the local level. For example, our results indicate that accessibility to educational and retail amenities displays a positive association with human mobility at the global level; yet, wide geographic variability exists at the local level. Accessibility to education amenities exerts a strong impact on promoting mobility in a few central and northern neighborhoods of Barcelona, but its effect is negligible in other neighborhoods. Similarly, accessibility to retail amenities is prominent shaping mobility in coastal central neighborhoods but relatively unimportant in other areas across Barcelona.

Additionally, we also presented evidence on the urban amenities linked to reduced human mobility levels. Our results, for example, revealed that differences in accessibility to professional, health and public transport amenities are negatively associated with human mobility in specific neighborhoods. These results suggest that greater accessibility to these amenities in these neighborhoods tend to deter mobility to neighborhoods with significantly lower accessibility levels for these services, and \textit{vice versa}.

Together, our findings contribute to inform spatially targeted urban policy interventions: first, by providing evidence on the extent of geographic variability in the relationship between urban amenities and human mobility; and, second, by identifying the set of neighborhoods and urban amenities associated with high and low mobility levels, or resident population retention rates. Ultimately our findings can effectively inform urban policy interventions seeking to reduce local levels of mobility and move closer towards the realization of a 15-minutes city or polycentric city by identifying the specific set of urban amenities requiring intervention in each neighborhood.

% key insights
To exemplify how local mobility models help to understand urban phenomena, we compared how the found patterns by our model differ when analyzing a current problem that affects Barcelona and many other cities in Europe: the rise of rental prices, which are linked to over-tourism and gentrification processes. One of the key novel insights we obtained is that {\em better access} to tourist attractions can push rental prices upwards as easily as {\em having} tourist attraction. A plausible explanation is that the presence of tourists drives the price down, but closeness to the attraction is still valuable. 

% limitations
Our work is scoped by the market share of the mobile phone operator, which is likely to be biased toward specific socio-economic and demographic groups. Given that we used aggregated, anonymized data, we cannot measure and therefore correct any biases emerging from the data. Since the usage of mobile phone data in data-for-good projects is an active area of research~\cite{oliver2020mobile}, we expect that new methods and other datasets will enable mitigating these biases.

% future work
We devise four main lines of future work: 
to integrate the several types of visitors in a common framework, to include other variables into the model, the definition of time-aware metrics, and multi-city analysis.
First, we focused on the origin and destination patterns of Barcelona residents. However, the city's floating population has a non-negligible share of commuters and tourists that should be taken into account in future iterations of the work. This line presents challenges as those visitors do not have an origin to map directly into the model.
Second, even though we have included population and demographic variables into our model, there may be a need for other variables. 
We focused on the importance of quantity and diversity of local amenities, but there may be other qualities that are relevant; if there are multiple 15-minutes cities, it is likely that some of them present better amenities than others, thus, attracting visitors. The availability of user-generated content, such as reviews and check-ins, may provide additional insights to deepen the analysis on the importance of urban amenities shaping these patterns of mobility.
Additionally, urban disciplines such as transportation may need to incorporate factors such as employment density, fine-grained population data, distribution of mode of transportation usage, among others. Some of the factors under study address these variables indirectly, such as the population counts and the socio-economic characterization, however, for urban planning practice they may need to be explicit.
Third, incorporating time into the model would enable the measurement of effects of public policy changes and urban interventions. There are emerging methods that extend GWR with temporal awareness~\cite{gmd-2019-292} that could be used in future analyses.
Finally, to advance on the path to creating inclusive and sustainable cities, quantitative methods combined with fine-grained datasets are required to understand and compare different cities. This will allow researchers and urban planners to distinguish between specific results from a case study and systematic subdivisions in cities that need to be made visible and measurable.

\section*{Acknowledgments}
This project has received funding from the European Union's Hori\-zon 2020 research and innovation programme under grant agree\-ment No. 857191 (IoTwins project). The funders had no role in study design, data collection and analysis, decision to publish, or preparation of the manuscript. Eduardo Graells-Garrido was partially funded by ANID Fondecyt de Iniciaci\'on 11180913.
We acknowledge the following software libraries used in the analysis: 
numpy~\cite{harris2020array}, scipy~\cite{virtanen2020scipy},
matplotlib~\cite{hunter2007matplotlib}, seaborn, 
PySAL~\cite{rey2010pysal}, mgwr~\cite{oshan2019mgwr}, pandas~\cite{mckinney2011pandas}, and geopandas.
Map data used in the accessibility analysis includes copyrighted data from OpenStreetMap contributors.
We thank the anonymous reviewers for their suggestions to improve this paper.
Finally, we thank the people from \emph{Ajuntament de Barcelona} and Vodafone for providing access to the data and for useful discussions.

\nolinenumbers

% Either type in your references using
% \begin{thebibliography}{}
% Text
% \end{thebibliography}
%
% or
%
% Compile your BiBTeX database using our plos2015.bst
% style file and paste the contents of your .bbl file
% here. See http://journals.plos.org/plosone/s/latex for 
%\bibliographystyle{plos2015}
%\bibliography{sample-base}

\end{document}